# Pressure Dependence of Energetic (≤160 keV) Focused Electron Beams Arising From Heated or Cooled ($LiNbO_3$) Pyroelectric Crystals

James D. Brownridge[a)]
*Department of Physics, Applied Physics and Astronomy, State University of New York at Binghamton, P.O. Box 6000, Binghamton, New York 13902-6000*

Stephen M. Shafroth[b)]
*Department of Physics and Astronomy, University of North Carolina at Chapel Hill, Chapel Hill, North Carolina 27599-3255*

**Abstract.** A new effect, "gas amplification of electron energy" is reported here; namely when a cylindrical pyroelectric crystal such as ($LiNbO_3$) is contained in a concentric cylindrical chamber and is heated and then allowed to cool in a dilute gas the maximum energy of the resultant focused electron beam more than doubles as the pressure increases from 0.05 to 4 mTorr for seven different gases.

Even though pyroelectric crystals have been known since ancient Greek times[1], surprising new effects and practical applications are constantly being discovered[1-12], for example; electrons accelerated by heating and cooling such crystals in dilute gases have produced energetic brehmsstralung and characteristic K x-rays up to Au.[8] There is a voluminous literature on electron emission by pyroelectric crystals. Reference 10, a comprehensive review article, is representative but does not suggest that such crystals can produce focused, energetic electron beams. Much of the current industrial use of these crystals is in infrared detection, sensitive temperature-change detectors and photonics, e.g. the emission of THz radiation in $LiNbO_3$ crystals[11]. For these reasons few studies of the behavior of these crystals in dilute gases have previously been done so no one has yet discovered the phenomena reported here, namely that if a $LiNbO_3$ crystal is heated and then allowed to cool in a dilute gas, the electron beam energy and intensity depends strongly on the gas pressure and less on gas type.

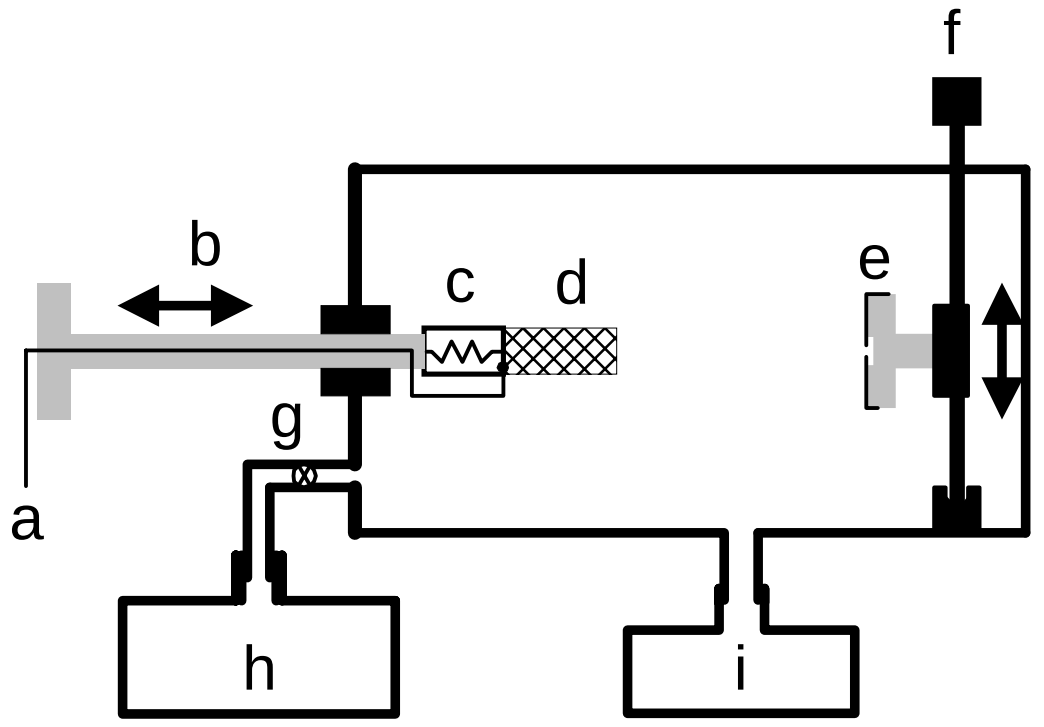

Fig. 1. Experimental arrangement. (a) Thermocouple and heater wires. (b) Axial motion of crystal. (c) 62 ohm resistor. (d) Cylindrical $LiNbO_3$ crystal. (e) 0.1 mm slit and surface barrier detector. (f) Motor, which moves the slit-detector system vertically. (g) Valve. (h) Gas container (i) Pump.

There are interesting practical applications of pyroelectric crystals in dilute gases for laboratory teaching[4] of X-ray and electron physics. The various parameters such as pressure, temperature etc. can be controlled by student-written LabView programs[5]. A further application of this work is to use a pyroelectric crystal to produce bremsstrahlung X-rays, which can fluoresce high-Z targets and produce very background-free characteristic K X-ray spectra[9] of elements at least up to Pb. The experimental arrangement used for this work is diagrammed schematically in Fig. 1. It was designed expressly to study the behavior of the focused electron beams, Vs pressure and gas type. Since results are geometry dependent, we used a cylindrical $LiNbO_3$ crystal, 10 mm x 4 mm dia. whose + z base was epoxied to a 62 ohm resistor. This assembly, which could be moved along the axis, was mounted axially in a glass cylinder 45 cm x 7.5 cm dia. The chamber was capped by metal end plates. The crystal is heated to ~180 $^{o}$C by passing a current (~120 mA) through the resistor. The crystal then cools naturally. The electrons are detected by a fully depleted 100 μm surface barrier detector located behind a 0.1 mm slit. Varying the distance between the slit and the crystal does not significantly effect the electron energy until the distance is ≤5 mm. The slit-detector system can be moved vertically across the beam (See ref. 7). These data were taken with the slit located at the focal point of the beam, i.e. ~22 mm. Typical electron spectra taken at 2 mTorr $N_2$ and different temperatures as the crystal cooled are shown in Fig. 2(a-g) The lowest energy peak at energy E is the true energy of the electrons. The peaks at 2E and 3E are due to pile up. (See ref. 6 for additional spectra). Figure 2(h and i) show the result of reducing the pressure from 2 mtorr to 0.2 mtorr. The electron energy drops by almost a factor of two. Figure 3 shows the electron counting rate (on a log scale) Vs. time. It also shows the cooling curve for the crystal and the pressure in the chamber. It can be seen from the inset that at 2470 sec the pressure is reduce over a period of ~ 30 sec by closing the leak valve to the gas supply which causes the electron count rate to increase dramatically and then fall to below its former value while simultaneously the beam is also defocused. This change in electron count rate and electron energy as the pressure is reduced is also shown in fig. 2(g-i).

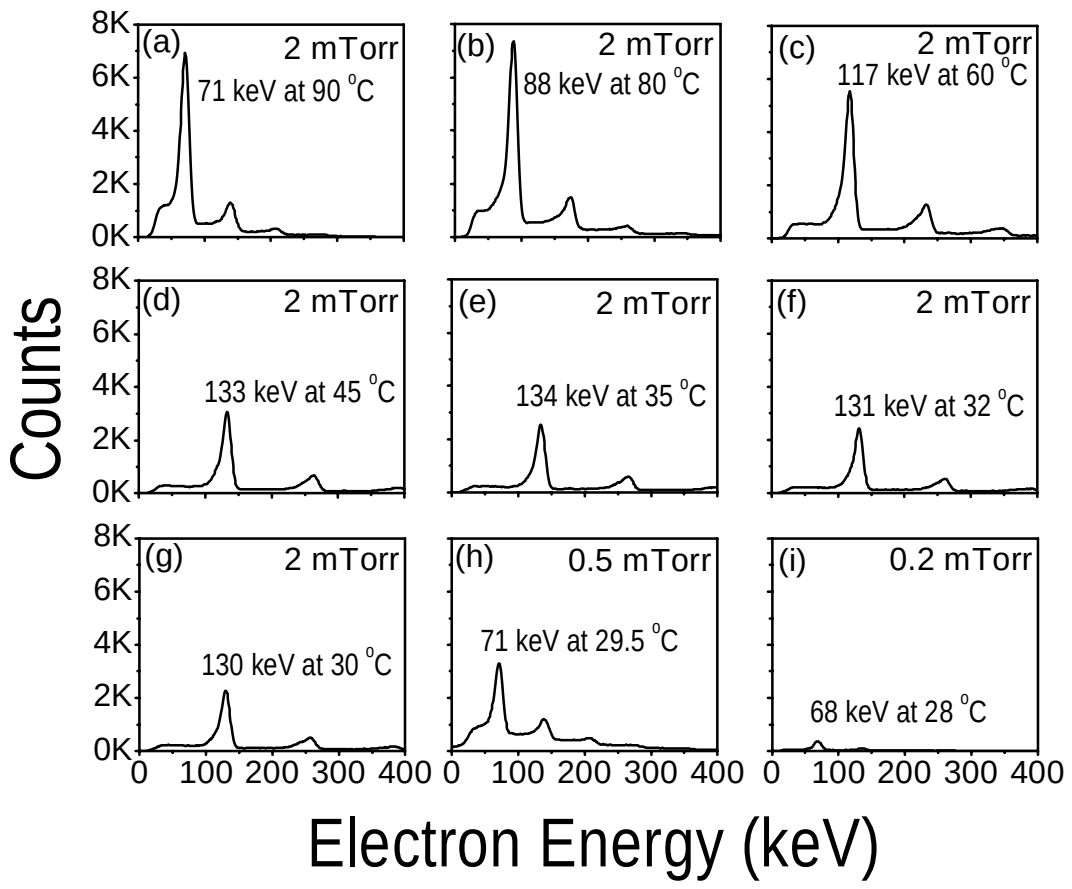

Fig. 2. Electron spectra at different temperatures and pressures as the crystal cooled. The pressure reduction began between (g) and (h) and continued through (i).

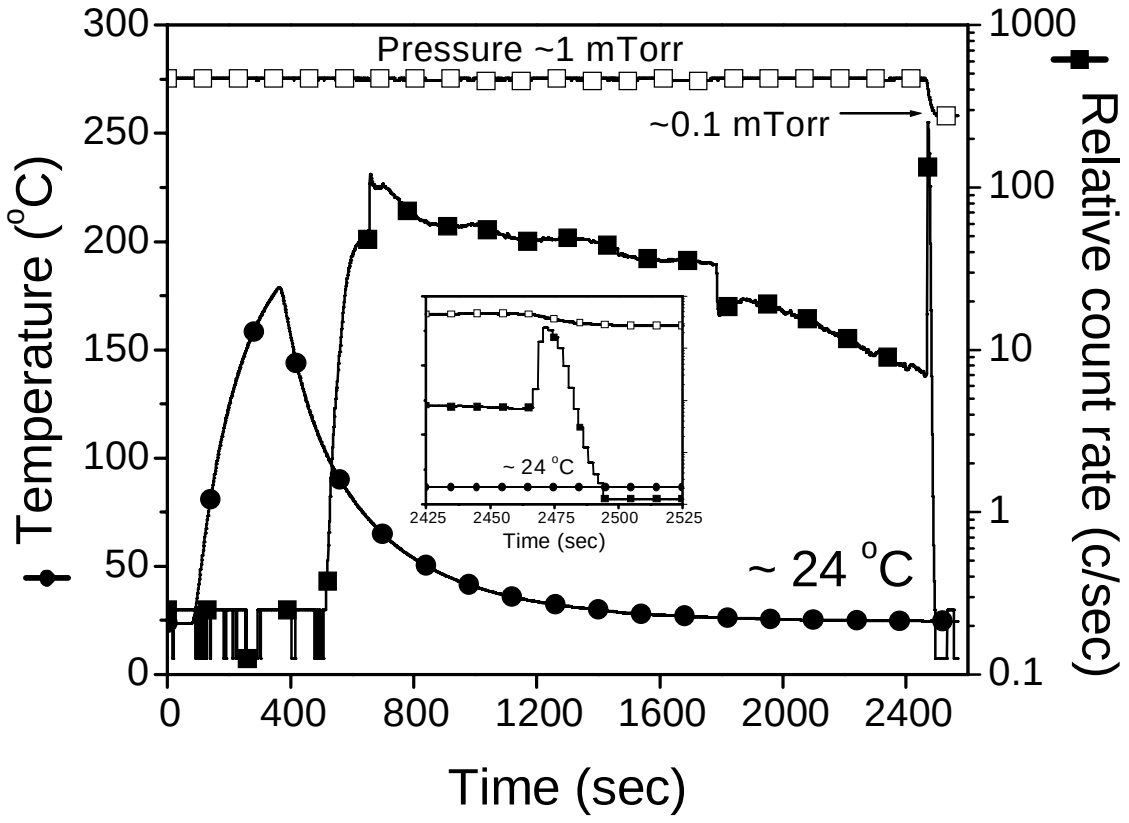

Fig. 3. Crystal temperature as indicated by thermocouple, solid circles, electron counting rate, solid squares, (log scale) and pressure, open squares. Inset: Log of electron count rate and gas pressure when the gas inlet valve was closed.

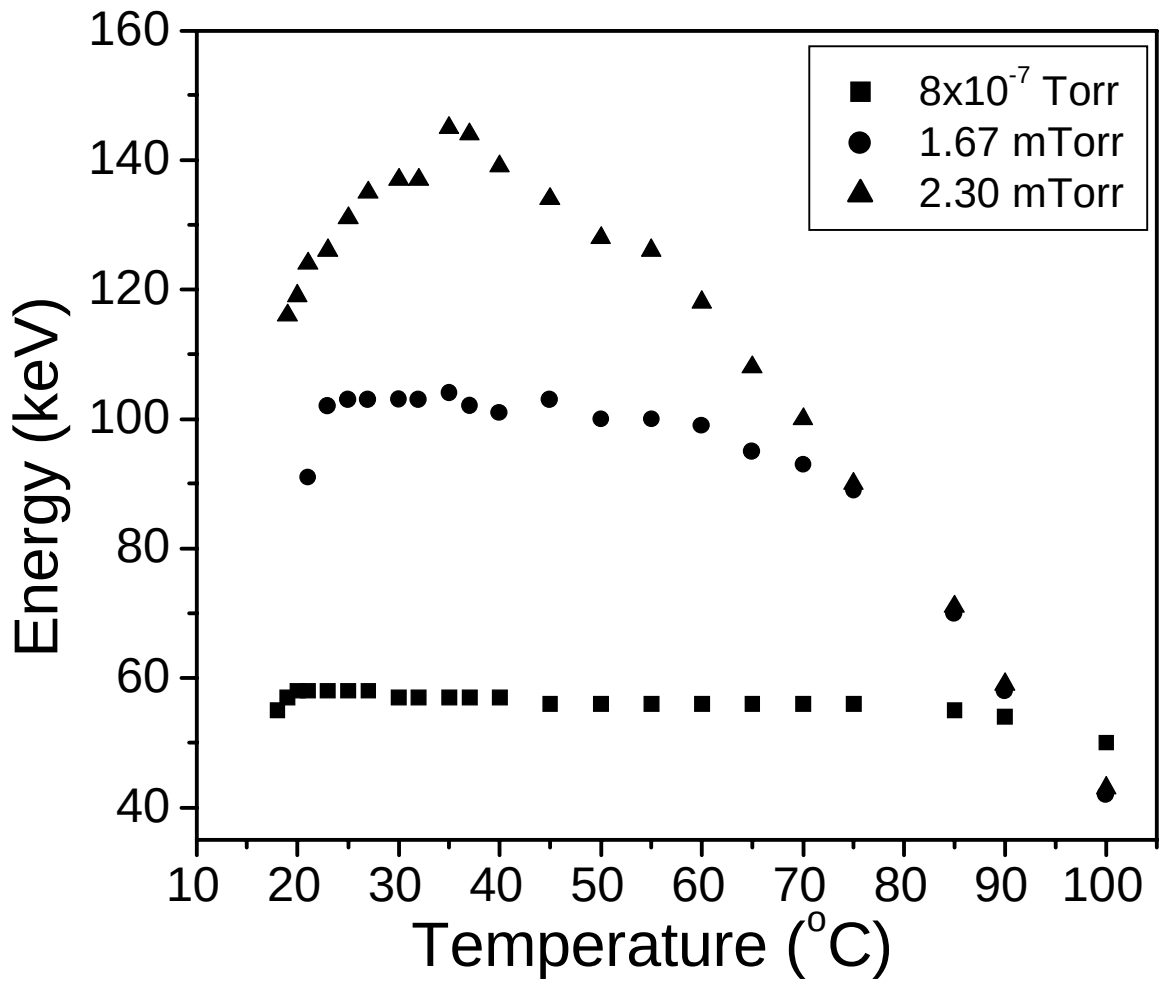

Fig. 4. Electron energy Vs crystal temperature at 3 different pressures as the crystal cools.

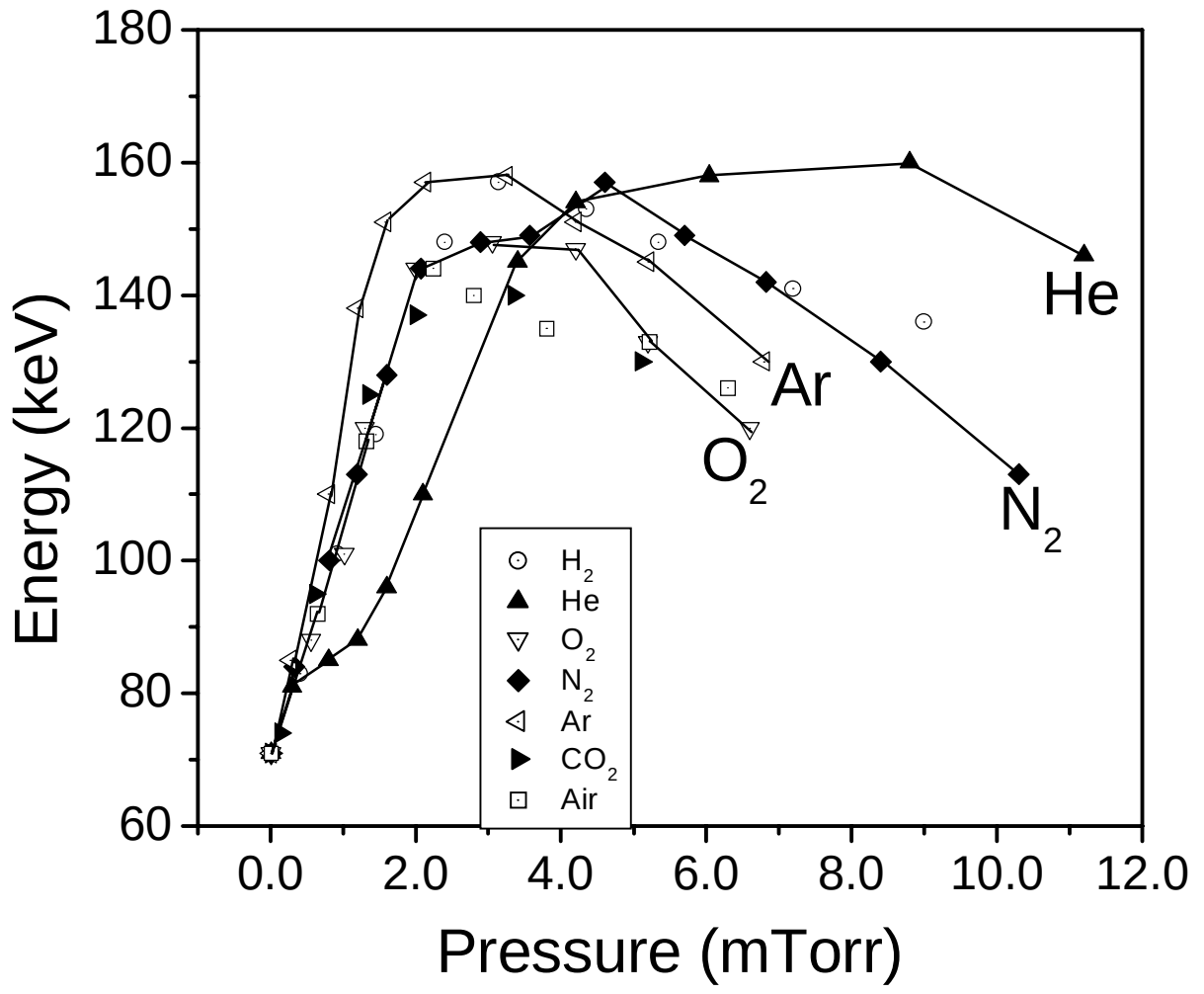

Fig. 5. Maximum electron energy Vs pressure for 7 gases. Each data point is the maximum electron energy observed as the crystal cooled from ~180 °C at the pressure indicated.

Figure 4 shows the electron energy as a function of temperature, which is directly related to cooling time (see fig 3) at three different pressures. At the lowest pressure ($8 \times 10^{-7}$ Torr) the electron energy slowly increases as the crystal cools. In this case the electron energy is more closely proportional to the exposed surface polarization charge[3] density. If the pressure is increased to 1.67 mTorr the electron energy slowly increases from ~50 keV to ~ 100 keV as the crystal cools from $90^0$ C to $25^0$ C. This is the gas amplification effect. We believe it is related to the formation of plasma at the crystal surface. Finally at 2.3 mTorr the electron energy increases further until the crystal cools to about 30 °C and then starts to drop with further cooling. Again this is the result of the net charge on the crystal surface (polarization and adsorbed ions) as well as the plasma at the surface.

Figure 5 summarizes the main effect," gas amplification of electron energy", reported in this paper, which as far as we know has not previously been reported in the literature. Data points were collected after adjusting the pressure and then cycling the crystal once completely over a thermal cycle. It shows that the maximum electron energy more than doubles in the range from 0.05 to 4 mTorr for a variety of gases. The maximum electron energy then levels off and decreases with further pressure increase. Also the rate of increase in energy with pressure starts out the same for most gases (~50 keV/mTorr) whereas in the case of He it is ~ 25 keV/mTorr starting at 1.5 mtorr, and in the case of Ar it is 57 keV/mTorr.

We are most grateful to our colleagues, Sol Raboy, Tom Clegg and Eugen Merzbacher for insightful discussions, and continued encouragement.

[a)] e-mail jdbjdb@binghamton.edu

[b)] e-mail shafroth@physics.unc.edu